%Paper: hep-th/9212019
%From: "David Lowe" <lowe@puhep1.Princeton.EDU>
%Date: Wed, 2 Dec 92 20:04:11 EST
%Date (revised): Thu, 3 Dec 92 12:42:41 EST

\input harvmac.tex
\input tables
\def\dpsi{\Delta_{\psi}}
\def\dphi{\Delta_{\phi}}

\Title{\vbox{\baselineskip12pt\hbox{PUPT-1362}
\hbox{hep-th/9212019}}}
{\vbox{\centerline{Conformal Models of}
\centerline{Two--Dimensional Turbulence}}}

\centerline{David A. Lowe \foot{lowe@puhep1.princeton.edu} }
\centerline{\it Joseph Henry Laboratories}
\centerline{\it Princeton University}
\centerline{\it Princeton, NJ 08544}
\bigskip
\noindent
Polyakov recently showed how to use conformal field theory to
describe two-dimensional turbulence. Here we construct an
infinite hierarchy of solutions,
both for
the constant enstrophy flux cascade, and the constant energy
flux cascade. We conclude with some speculations concerning
the stability and physical meaning of these solutions.

\Date{November, 1992}
%\draft
\noblackbox

\newsec{Introduction}

Recently, Polyakov proposed a method for using conformal field
theory to describe time averaged probability distributions in
two-dimensional turbulence
\nref\sasha{A.M. Polyakov, Princeton preprint PUPT-1341.}%
\refs{\sasha}.
The essential idea was to use
a local quantum field theory to describe correlation functions
in the inertial range. One expects physics to be scale invariant
in this range, so automatically the theory becomes conformally
invariant
\nref\pol{A.M.~Polyakov, {\it JETP Lett.} {\bf 12} (1970) 381.}
\refs{\pol}.

Polyakov formulated the Navier-Stokes equations in terms of operators
in the CFT, and used this to derive a matching condition that
the effective CFT must satisfy. This followed from imposing the
condition of constant enstrophy flux, which is the basis
for the usual Kolmogorov-type arguments in two-dimensions
\nref\krai{R. Kraichnan, {\it Phys. of Fluids} {\bf 10}, (1967) 1417.}
\refs{\krai}.
In \refs{\sasha} one solution of the equations was found,
namely the $(2,21)$ minimal model.

In this paper we construct an infinite series of solutions
to the matching condition. In addition, solutions
corresponding to a constant flux of energy are considered.
This is known as the reverse cascade, since as shown in \refs{\krai}
energy flows from small scales to large scales. Again an
infinite hierarchy of solutions seems to exist.
We conclude with some speculations about the stability and physical
meaning of these models.

\newsec{Conformal field theory of turbulence}

The Navier-Stokes equation in two-dimensions takes the form
\eqn\navier{
\dot \omega + \epsilon_{\alpha \beta} \partial_{\alpha} \psi
\partial_{\beta} \partial^2 \psi = \nu \partial^2 \omega~,
}
where $\omega$ is the vorticity ($\omega = \partial^2 \psi$) and
$\psi$ is the stream function, related to the velocity
by $v_{\alpha} = -\epsilon_{\alpha \beta} \psi$. Here $\nu$ is
the viscosity.

Polyakov's proposal is that correlators such as
\eqn\exam{
\vev{ \omega(q_1) \cdots \omega(q_n)} ~,
}
are described by conformal field theory when all the
$q_i$ and all sums over partitions of the $q_i$ are in the
inertial range: $1/R \ll q \ll 1/a$. Here $R$ is the
infrared cutoff, and $a$ the ultraviolet cutoff, which
is closely related to the viscosity. When considering correlators
in position space, one should think, for example, of
the stream function being decomposed as
\eqn\strm{
\psi=\psi_c+\psi_{IR}~,
}
where $\psi_c$ is the conformal part, and $\psi_{IR}$ is a piece that
fluctuates only at IR scales.

Equation \navier\ is used to define the operator $\dot \omega$. For
this to be well-defined, a point splitting regularization must be used.
The result is that
\eqn\omdot{
\dot \omega \sim |a|^{2\dphi- 4\dpsi} ( L_{-2} \bar L_{-1}^2
- L_{-1}^2 \bar L_{-2} ) \phi~,
}
where $\phi$ is the minimal dimension operator that appears on the
RHS of
\eqn\ope{
\psi \times \psi \sim \phi + \cdots~.
}

As we take the viscosity to zero, $a\to 0$, so we should require
that $\dot \omega$ vanish in order that the inviscid Hopf
equation be satisfied, i.e.
\eqn\hopf{
\vev{\dot \omega(\vec x_1) \omega(\vec x_2) \cdots
\omega(\vec x_n) } + \vev{  \omega(\vec x_1) \dot  \omega(\vec x_2)
\cdots \omega(\vec x_n) } + \cdots = 0 ~.
}
This requires the condition
\eqn\ineone{
\dphi > 2\dpsi~.
}

The condition that the effective conformal field theory match
correctly with the enstrophy input at large scales and
dissipated at small scales is that the enstrophy flux is constant.
This is written as
\eqn\match{
\vev{ \dot \omega(r) \omega(0) } = {\rm const}~,
}
i.e. is independent of $R$ and $r$. Inserting the conformal
fields one finds zero contribution. Polyakov then
conjectures that the contributions arising from the
IR part of the correlator must have a piece that scales
with $R$ in the way required by the anomalous dimensions
of the conformal fields.
We have then that
\eqn\mat{
\vev{ \dot \omega(r) \omega(0) } \sim |a|^{2\dphi-4\dpsi} R^{-2\dphi
-2\dpsi-6} r^0~.
}
Demanding this be independent of $R$ leads to the condition
\eqn\enst{
\dpsi+\dphi = -3~.
}
Together with \ineone\ this implies $\dpsi<-1$ so the
energy spectrum is always steeper than the Kolmogorov-Kraichnan
result of $E(k) \sim k^{-3}$. Here one finds
$E(k) \sim k^{4\dpsi +1}$.

In the following we will explore solutions of these equations,
and also consider the case of constant energy flux, which
may lead to an inertial range with a cascade of energy
from small to large scales.

\newsec{Minimal model solutions for constant enstrophy flux}

Although the solutions of these equations for any conformal field theory
are of interest, the minimal model solutions are a particularly
interesting class, since only then are there a finite number of
operators with negative dimension.

Before proceeding, let us review a few relevant facts about
the minimal models
\nref\bpz{A.A.~Belavin,A.M.~ Polyakov, A.B.~Zamolodchikov, {\it Nuc. Phys. }
{\bf B241} (1984) 333.}
\refs{\bpz} .
The $(p,q)$ minimal model (where $p<q$, with $p$
and $q$ relatively prime) contains $\half (p-1)(q-1)$
degenerate primary operators
which we label as $\psi_{m,n}$, where $1\leq n < q$ and
$1\leq m <p$. These have conformal dimensions
\eqn\dimens{
\Delta_{m,n} = { {(pn-qm)^2-(p-q)^2 } \over {4pq}}~,
}
and satisfy the fusion rules
\eqn\fuse{
\eqalign{
\psi_{m_1, n_1} \times \psi_{m_2, n_2} =
& \sum_{i=|m_1-m_2|+1}^{{\rm Min}(m_1+m_2-1, 2p-m_1-m_2-1) }\cr
& \sum_{j=|n_1-n_2|+1}^{{\rm Min}(n_1+n_2-1, 2q-n_1-n_2-1) }
D_{(m_1,n_1) (m_2,n_2)}^{ (i,j)} \psi_{i,j} ~,\cr
}}
where $i$ runs over odd numbers if $m_1+m_2-1$ is odd,
or even numbers if $m_1+m_2-1$ is even, and similarly for $j$.
The $D_{(m_1,n_1) (m_2,n_2)}^{ (i,j)}$ are the structure constants
of the conformal field theory. Note there is a symmetry of the
conformal dimensions $\Delta_{m,n} = \Delta_{p-m,q-n}$.

\subsec{Solutions with $\psi=\psi_{1,s}$ }

Let us begin by classifying the solutions when $\psi=\psi_{1,s}$.
Let $\phi=\psi_{1,t}$, and parametrize $q$ by
\eqn\qform{
q= t p+n ~.}
Equation \enst\ then gives
\eqn\peqn{
p = {{2n(8-s-t) }\over {2 -s^2-16 t + 2 s t+t^2}} ~.}

In order that, $\phi$ be the minimal dimension operator on the RHS of \ope\ we
must have either
\item{(a)} $p~ {\rm Min}(2s-1,2q-2s-1) < q$, or
\item{(b)} $|n| <p $.

Consider case $(a)$.
Set $t=2s-1$, since $t=2q-2s-1$ is simply obtained by using invariance
under $s\to q-s$.
Substituting this
into \peqn\ we obtain
\eqn\neqn{
n={ {(19-38s+7s^2) p } \over {9-3s}} ~.}
However for $s>4$ this means $n<0$ which is not possible, so it is only
necessary to look for solutions with $s \in [1,4]$. The only solution is
the $(2,21)$ model with $\psi=\psi_{1,4}$ and $\phi=\psi_{1,7}$ as found
in
\refs{\sasha}.

Now consider case $(b)$, $|n|<p$. The condition $\dpsi<-1$ implies that
$\dphi>-2$, so
\eqn\tinequal{
{{ (pt-q)^2-(p-q)^2} \over {4pq}} > -2~, }
which leads to
\eqn\tineq{
p(12 t -t^2-11) + 10-2t>0 ~.}
Since the fusion rules \fuse\ imply $t $ is an odd number, there are no
solutions of this inequality for $t>9$. It remains then to find the
solutions for $t \leq 9$. There are a finite number of these as may be
seen from the inequality derived from $|n|<p$,
\eqn\sineqal{
\biggl|{n \over p}\biggr| =
\biggl| {{2-s^2-16 t+ 2s t+t^2}\over {2(8-s-t)}} \biggr|
<1 ~.}
These solutions are displayed in appendix A, along
with all other solutions for $q<500$.
Note that similar
manipulations may be carried out whenever $\psi=\psi_{m,n}$ and for
fixed $m$ there will be a finite number of solutions. However $m$ is
unbounded so in general one expects an infinite number of solutions.
This will be confirmed in the following.
It should also be noted that the results of
\nref\frisch{P.L.~Sulem and U. Frisch, {\it J. Fluid Mech.}
{\bf 72} part 3 (1975) 417.}
\refs{\frisch} tell us that solutions with spectra
steeper than $k^{-4}$ correspond to vanishing flux of enstrophy.

One situation in which $\phi$ need not be the minimal dimension
operator on the RHS of \ope\ is when $\phi$ is degenerate on level
2, causing its contribution to $\dot \omega$ to vanish. The leading
contribution to $\dot \omega$ would then come from the next to
minimal operator on the RHS of \ope\ . It turns out, however, that
this situation never arises in minimal model solutions.

\subsec{Parametric solution of the general equations}

For an arbitrary $(p,q)$ minimal model solution with
$\psi=\psi_{m,n}$ and $\phi=\psi_{s,t}$, equation \enst\ may be
rewritten
\eqn\ea{
r^2+l^2 -2(p^2+q^2-8pq) =0 ~,
}
where $l= (n p -q m)$ and $r= (t p-q s)$. Defining $y=q-4p$ this may be
diagonalized
\eqn\eb{
r^2+l^2 +30p^2 - 2y^2 =0~.
}
This is a homogeneous quadratic
Diophantine equation in four variables. It is well
known
\nref\dioph{Dickson,``History of the theory of numbers, vol.II,
Diophantine analysis,''pub. Carnegie Institute, Washington,1920}%
\refs{\dioph}
that the general solution to this equation in integers
may be written as
\eqn\param{
\eqalign{
r &= 2a^2 -30 b^2-c^2 \cr
l &= r+c(2c-4a) \cr
p &=|b(2c-4a)| \cr
y &=|r+a(2c-4a)| ~,\cr
}
}
where $a,b$ and $c$ are integer parameters, and solutions differing by a
constant are to be identified. Unfortunately it is difficult to
implement the conditions $\dpsi<-1$ and $\dphi$ is minimal directly
on this parametric solution, so many spurious solutions are generated.
It is, however, a useful step in reducing the problem.

\subsec{An infinite series of solutions}

One interesting infinite sequence of solutions that do solve the
constraints imposed by the OPE is the following:
\eqn\infin{
\eqalign{
\phi&= \psi_{1,9} \cr
q&= 9p \pm 1 \cr
l+(2p \pm 1) \sqrt{5} &= (9\pm 4\sqrt{5})^j (1\pm \sqrt{5}) \cr
l &= np-mq ~,\cr
}
}
where $\psi=\psi_{m,n}$, and $j$ is an
integer parameter. Note that  $\psi$ is uniquely determined by solving
the linear Diophantine equation for $n$ and $m$. These solutions all
satisfy
\eqn\dph{
\dphi = {{1-(p-q)^2}\over {4pq}}~,
}
so $\phi$ is the minimal dimension operator in the model. In addition
it may be seen that $\phi$ appears in the OPE $\psi \times \psi$ for
$j>1$. As $p$ and $q$ become large, the critical exponent
$4\dpsi +1 \to -3.89$~.

\subsec{Inequalities for general solutions}

A stringent upper bound on the ratio $p/q$ comes from
equation \ea\ . Subtracting off $r^2+l^2$ one obtains
\eqn\ineqp{
-2(p^2+q^2- 8pq) \leq 0 ~,
}
so that
\eqn\pqineq{
{p\over q} < 4-\sqrt{15}~.
}

One may also show that the $(2,21)$ model is the only model that
does not satisfy $|r|<p$. This is done as follows. Suppose $|r| \geq p$.
In order that $\phi$ be the minimal dimension operator that appears
on the RHS of \ope\ we must have that $\phi=\psi_{1,t}$ and that
$r= p t -q$ satisfy
\eqn\ineb{
p\leq r < q~.
}
The parameter $t$ will take its
maximal value so that $|r|$ is minimized. For $\psi=\psi_{m,n}$ we have
therefore that $t=2n-1$. Equation \ineb\ leads to
\eqn\inet{
t<{{2q}\over {p}}~,
} so that
\eqn\inen{ n<{q\over p} +\half~.}
Now the condition $\dpsi < -1$ can be rewritten as
\eqn\innen{
m < \sqrt{ (p/q)^2 +1 - 6(p/q) } + 1 + {p\over {2q}} ~.
}
Using \pqineq\ one finds that $m=1$ is the only possibility. Since
all models with $m=1$ were classified above, one finds that the
$(2,21)$ model is the only one that satisfies $|r|\geq p$.

This fact may be used to place a lower bound on $p/q$ which is satisfied
for all models other than the $(2,21)$ model. The inequality $\dpsi<-1$
implies that $\dphi>-2$ so that \enst\ is satisfied. Using $|r|<p$ tells
us
\eqn\ined{
{ {p^2-(p-q)^2}\over {4pq}} > -2~,
}
which implies that
\eqn\inee{
{p\over q} > {1\over {10}}~,
}
for all solutions except the $(2,21)$ model.

These bounds on $p/q$ then imply that the central charge
\eqn\cent{
c= 1 - {{6(p-q)^2} \over {pq}}~,
}
is minimized for the $(2,21)$ model, where $c=-50{4\over 7}$.
Unfortunately, the Zamolodchikov c-theorem
\nref\zam{A.B.~Zamolodchikov, {\it JETP Lett.} {\bf 43}
(1986) 730.}
\refs{\zam}
does not apply to
conformal theories that do not satisfy reflection-positivity
(which is always the case here). Otherwise one would then be able to
argue that the $(2,21)$ model is at a fixed point under perturbations
of the conformal field theory.
The minimal value of $c$ for the $(2,21)$ model does mean this model
is stable under a large class of perturbations. Perhaps these
are particularly relevant for hydrodynamic stability.

\subsec{Parity properties}

All $(p,q)$ minimal models possess a $Z_2$ symmetry
when one of $p$ and $q$ is odd, the other even.
If one considers the
whole Kac table $\psi_{m,n}$, with $1\leq m <p$ and $1\leq n <q$, and
restricts consideration to operators with $n+m$ even, then
operators with $n$ and $m$ odd are in the odd parity sector, and
operators with $n$ and $m$ even are in the even parity sector.

We normally think of $\psi$ as a pseudoscalar operator, so this should
be in the odd parity sector. On the other hand, we know that
$\phi=\psi_{t,s}$ has $s$ and $t$ both odd, so must also be in the
odd parity sector. This leads one to conclude that there is no
well defined parity for these solutions. In fact, $\psi$ cannot be
in the even parity sector either. This follows from the fact that
$\dpsi + \dphi$ then equals an odd number divided by an even number,
so cannot satisfy \enst\ .
It seems then that the minimal model solutions should break parity
invariance.
If correlation functions are symmetric under an additional $Z_2$
symmetry, this may then be used to define spatial parity, so in
some cases, parity preserving models may exist.

\newsec{Constant energy flux cascade}

The constant energy flux inertial range may be considered in the same
way as above. The Navier--Stokes equations are
\eqn\nav{
\dot v_{\alpha} + v_{\beta} \partial_{\beta} v_{\alpha} =
-{1\over {\rho}}\partial_{\alpha} p + \nu \partial^2 v_{\alpha}~.
}
Taking the divergence of this we get
\eqn\press{
{1\over {\rho}}\partial^2 p = - \partial_{\alpha} \partial_{\beta} (
v_{\alpha} v_{\beta} )~,
}
determining the pressure in terms of the velocity field.
Neglecting viscosity, and working in momentum space we find that
\eqn\vdot{
\dot v_{\alpha} (q) = - (v_{\beta} \partial_{\beta} v_{\alpha} )(q) +
 { {q_{\alpha} q_{\beta}} \over {q^2}} (v_{\gamma} \partial_{\gamma}
v_{\beta} ) (q)~.
}
As before, we define $v_{\beta} \partial_{\beta} v_{\alpha}$ using
a point--split regularization, and use the operator product
expansion to obtain
\eqn\vope{
\eqalign{
v_{\beta} \partial_{\beta} v_z(\vec x)& \sim
|a|^{2\dphi-4\dpsi-2} B L_{-1} \phi + |a|^{2\dphi- 4\dpsi} (
L_{-2} {\bar L}_{-1} + A L_{-1}^2 {\bar L}_{-1} ) \phi +\cdots \cr
v_{\beta} \partial_{\beta} v_{\bar z}(\vec x)& \sim
|a|^{2\dphi-4\dpsi-2} B {\bar L}_{-1} \phi + |a|^{2\dphi- 4\dpsi} (
{\bar L}_{-2}  L_{-1} + A {\bar L}_{-1}^2  L_{-1} ) \phi +\cdots \cr
}}
where $A$ and $B$ are constants, determined by the operator product
expansion. Note that by taking the curl of \vope\ we find agreement
with \omdot\ as desired. Inserting \vope\ into \vdot\ we find that
the piece proportional to $B$ drops out, and so
\eqn\vdott{
\eqalign{
\dot v_z(q) &\sim |a|^{2\dphi-4\dpsi}\biggl(
-\bigl( ( L_{-2} {\bar L}_{-1} +
A L_{-1}^2 {\bar L}_{-1} ) \phi\bigr) (q) \cr
&\quad + {{q_z}\over {q^2}} \biggl(
q_z \bigl( ({\bar L}_{-2}  L_{-1} +
A {\bar L}_{-1}^2  L_{-1} ) \phi \bigr)(q) +
q_{\bar z} \bigl( ( L_{-2} {\bar L}_{-1} +
A L_{-1}^2 {\bar L}_{-1} ) \phi \bigr) (q)
\biggr) \biggr) ~.\cr
}
}
Now, following the same kind of arguments as lead to \enst\ , the
condition of constant energy flux
\eqn\enflux{
\vev{ \dot v_{\alpha}(r) v_{\alpha}(0) } \sim r^0~,
}
leads to the condition
\eqn\energ{
\dpsi + \dphi = -2~.
}
The inequality $\dphi > 2 \dpsi$, obtained by requiring the
vanishing of \vdott\ as $a\to 0$ means that
\eqn\eineq{
\dpsi < -2/3, \qquad \qquad \dphi> -4/3~, }
so the energy spectrum must be steeper than the Kolmogorov value.

\newsec{Minimal model solutions for constant energy flux}

Now, let us proceed to find minimal models that solve these
conditions.

\subsec{Solutions with $\psi=\psi_{1,s}$ }

The proof is the same as before. Let $\phi=\psi_{1,t}$ and
parametrize $q$ by $q=pt+n$. Equation \energ\ leads to
\eqn\npenerg{
{n\over p} = {{s^2-t^2 + 12 t -2 st-2}\over {2(t+s-6)}}~.
}
If $p~{\rm Min}(2s-1,2q-2s-1) < q$ then we must have that
$t=2s-1$ (provided we shift $s\to q-s$ as appropriate).
Substituting this into \npenerg\ one finds no solution for
integer $s$ with $n>0$.

The solutions must therefore satisfy $|n| <p$. The inequality
$\dphi>-4/3$ then gives
\eqn\abc{
-t^2+ 28/3 t -25/3 >0~.
}
Since $t$ is odd this means $t\leq 7$. Again this leads to
a finite number of solutions, which are displayed in
appendix B, along with all other solutions for $q<500$.
Note that solutions with spectra steeper than
$E(k) \sim k^{-8/3}$ must have vanishing energy flux,
as shown in \refs{\frisch} .

\subsec{Parametric solution of the general equation}

For the $(p,q)$ minimal model with $\psi=\psi_{m,n}$ and
$\phi=\psi_{s,t}$ equation \energ\ becomes
\eqn\ener{
r^2+l^2+16 p^2 - 2y^2 =0~,
}
where $l=np-qm$, $r=tp-sq$ and $y=q-3p$. The parametric solution
of this equation is then
\eqn\parma{
\eqalign{
r &= 2a^2 -16b^2 -c^2 \cr
l &= r + 2c(c-2a) \cr
p&= | 2b(c-2a)| \cr
y&= |r+ 2a(c-2a) | \cr
}
}
where $a,b,c$ are integer parameters and solutions differing by a
constant are to be identified. The additional conditions
imposed by the OPE are difficult to impose on this solution
so spurious solutions are still generated.

\subsec{Inequalities}

Equation \ener\ leads to the inequality
\eqn\ineqg{
-2(p^2+q^2-6pq) <0  ~,}
which means that
\eqn\pqbound{
{p\over q} < 3- \sqrt{8}~.
}

Following the same reasoning as before one may also prove that
all solutions must satisfy $|r|<p$. Suppose $|r|\geq p$. In order
that $\phi$ be minimal dimension we must have $|r|<q$. This
leads to $t< 2q/p$ and $n<q/p +\half$. Using the condition
$\dpsi< -2/3$ one finds that
\eqn\mineq{
m< \sqrt{1+ ({p\over q})^2 - {{14p} \over {3q}} } +1 + {p\over {2q}}~.
}
Together with \pqbound\ this is only satisfied for $m=1$. We have
already determined that all solutions with $m=1$ satisfy
$|r|<p$, so this must be true in general.

This fact may then be used to find a lower bound on $p/q$ by using
$\dphi>-4/3$ which gives
\eqn\pblower{
{{p^2-(p-q)^2 }\over {4pq}} > -{4\over 3}~,}
so that
\eqn\pblow{
{p\over q} > {3\over {22}} ~.}

\newsec{Stability?}

Probably some of these solutions do not correspond to
stable flow distributions. One would like to formulate
an additional criterion, in terms of the effective
conformal field theory description, that these solutions must satisfy.
One essential condition that must be satisfied is that of
reality of velocity correlation functions. This means velocity
correlators must satisfy a Cauchy-Schwarz inequality. Unfortunately,
it turns out to be impossible to neglect the IR sector of the
theory in this type relation. For example, if one works in momentum
space, one obtains relations such as
\eqn\cauchy{
\vev{ \omega(p_1) \omega(p_2)| \omega(p_3)}^2 =
\vev{\omega(p_1) \omega(p_2) | \omega(-p_1) \omega(-p_2) }
\vev{\omega(p_3)  | \omega(-p_3) }~,
}
which explicitly shows that a zero momentum intermediate state
appears.

Perhaps then, all (or at least some) of these solutions do correspond
to quasi-static equilibrium solutions. The picture that seems to emerge
is that of an infinite hierarchy of different scaling behaviors,
the stability of which depends strongly on the IR boundary conditions/
stirring forces.

There is actually some experimental evidence for this hypothesis.
It is well known that in two-dimensional turbulence coherent vortex
structures appear at the IR scale. The distribution of these
seems to depend on the type of stirring forces used as in
\nref\legras{ B. Legras, P. Santangelo and R. Benzi,
{\it Europhys. Lett.,} {\bf 5} (1), (1988) 37.}
\refs{\legras}.
There it was found that when the stirring forces caused
the coherent vortices to be unstable, the energy spectrum
scaled as $E(k) \sim k^{-3.5}$ or $k^{-3.6}$, very close to
the exact value for the $(2,21)$ model of $E(k)\sim k^{-25/7}$.
On the other hand, when the stirring forces favored coherent
vortex formation, a scaling law of $E(k)\sim k^{-4.2}$ was found
which happens to correspond closely to the third model shown in
appendix A, the $(3,25)$ model. It is also amusing to note
that results from a lower resolution simulation quoted in
\refs{\legras}
had a scaling law of $E(k)\sim k^{-4.7}$ which is close
to that of
the second model in appendix A.

It is natural to regard the $(2,21)$ model as the simplest
of the solutions to \enst\ since it has the lowest number of
degenerate primary operators. It is not too surprising, then, that it should
correspond to the simplest behavior of the fluid. As the
distribution of the coherent vortices becomes more complicated,
perhaps the whole hierarchy of models may be observed. Probably
there is some kind of entropy principle that must be taken into
account, which may restrict us, in practice, to observation
of only the first few scaling behaviors.

\newsec{Conclusion}

An infinite sequence of models have been found which
should describe turbulence in two dimensions. Perhaps
the physical solutions must satisfy some additional condition
which is necessary to correctly match the effective conformal
field theory with the correct IR and UV behavior. If not,
then an infinite hierarchy of inertial ranges seems likely,
although perhaps only the first few of these may be
observed in practice.

\bigskip
\centerline{\bf Acknowledgements}

The author wishes to thank  A.M. Polyakov for suggesting this
project, and for helpful discussions.
This research was supported in part by DOE grant DE-AC02-76WRO3072,
NSF grant PHY-9157482, and James S. McDonnell Foundation grant No.
91-48.

\vfill
\eject
\appendix{A}{Solutions of the constant enstrophy flux
condition for $q<500$}
\vskip .5cm
\begintable
 (p,q) | $\psi$ | $\phi$ | $4 \dpsi +1$ & \cr
 (2,21) | $\psi_{1,4}$ | $\psi_{1,7}$| $-25/7$&$\approx -3.57$ \nr
 (3,25) | $\psi_{1,11}$ | $\psi_{1,9}$ | $-23/5$&$\approx -4.60 $ \nr
 (3,26) | $\psi_{1,5}$ | $\psi_{1,9}$| $-55/13$&$ \approx -4.23$\nr
 (6,55) | $\psi_{1,14}$ | $\psi_{1,9}$| $-41/11$&$\approx -3.73$ \nr
 (7,62) | $\psi_{1,13}$ | $\psi_{1,9}$| $-125/31$&$\approx -4.03$ \nr
 (8,67) | $\psi_{3,28}$ | $\psi_{3,25}$| $-302/67$&$\approx -4.51$ \nr
 (9,71) | $\psi_{4,32}$ | $\psi_{7,55}$| $-173/35$&$\approx -4.99$ \nr
 (11,87) | $\psi_{2,16}$ | $\psi_{3,23}$| $-1605/319 $&$\approx -5.03$ \nr
 (11,91) | $\psi_{2,14}$ | $\psi_{3,25}$| $-355/77$&$\approx -4.61$ \nr
 (11,93) | $\psi_{2,20}$ | $\psi_{3,25}$| $-1515/341$&$\approx -4.44$ \nr
 (14,111) | $\psi_{1,8}$ | $\psi_{1,7}$| $-187/37$&$\approx -5.05$ \nr
 (14,115) | $\psi_{1,6}$ | $\psi_{1,9}$| $-109/23$&$\approx -4.74$ \nr
 (16,135) | $\psi_{7,56}$ | $\psi_{7,59}$| $-40/9$&$\approx -4.44$ \nr
 (21,166) | $\psi_{4,31}$ | $\psi_{7,55}$| $-2895/581$&$\approx -4.98$ \nr
 (22,179) | $\psi_{1,10}$ | $\psi_{1,9}$| $-865/179 $&$\approx -4.83$ \nr
 (25,197) | $\psi_{11,87}$ | $\psi_{9,71}$| $-4919/985$&$\approx -4.99$ \nr
 (26,205) | $\psi_{5,39}$ | $\psi_{9,71}$| $-2659/533$&$\approx -4.99$ \nr
 (26,213) | $\psi_{9,76}$ | $\psi_{5,41}$| $-4325/923$&$\approx -4.69$ \nr
 (23,217) | $\psi_{6,62}$ | $\psi_{9,85}$| $-2467/713$&$\approx -3.46$ \nr
 (26,223) | $\psi_{1,12}$ | $\psi_{1,9}$| $-965/223$&$\approx -4.33$ \nr
 (27,229) | $\psi_{10,88}$ | $\psi_{19,161}$| $-3025/687$&$\approx -4.40$ \nr
 (25,234) | $\psi_{9,79}$ | $\psi_{11,103}$| $-53/15$&$\approx -3.53$ \nr
 (32,267) | $\psi_{11,89}$ | $\psi_{3,25}$| $-1615/356$&$\approx -4.54$ \nr
 (35,277) | $\psi_{14,110}$ | $\psi_{11,87}$| $-9617/1939$&$\approx -4.96$
\endtable
\vskip .5cm
\begintable
 (p,q) | $\psi$ | $\phi$ | $4 \dpsi +1$ & \cr
 (29,280) | $\psi_{15,139}$ | $\psi_{3,29}$| $-94/29$&$\approx -3.24$ \nr
 (33,287) | $\psi_{14,118}$ | $\psi_{13,113}$| $-1889/451$&$\approx -4.19$ \nr
 (33,299) | $\psi_{10,86}$ | $\psi_{1,9}$| $-1145/299$&$\approx -3.83$ \nr
 (34,311) | $\psi_{15,142}$ | $\psi_{27,247}$| $-19793/5287$&$\approx
-3.74$ \nr
 (35,313) | $\psi_{6,58}$ | $\psi_{1,9}$| $-1235/313$&$\approx -3.95$ \nr
 (38,333) | $\psi_{13,110}$ | $\psi_{17,149}$| $-235/57$&$\approx -4.12$ \nr
 (34,335) | $\psi_{1,16}$ | $\psi_{1,9}$| $-209/67$&$\approx -3.12$ \nr
 (34,335) | $\psi_{15,154}$ | $\psi_{7,69}$| $-3469/1139$&$\approx -3.05$ \nr
 (39,346) | $\psi_{17,155}$ | $\psi_{31,275}$| $-9031/2249$&$\approx -4.02$ \nr
 (43,347) | $\psi_{11,89}$ | $\psi_{15,121}$| $-71719/14921$&$\approx
-4.81$ \nr
 (47,378) | $\psi_{15,119}$ | $\psi_{23,185}$| $-14311/2961$&$\approx
-4.83$ \nr
 (39,379) | $\psi_{14,142}$ | $\psi_{25,243}$| $-1205/379 $&$\approx -3.18$ \nr
 (44,397) | $\psi_{21,194}$ | $\psi_{1,9}$| $-1535/397$&$\approx -3.87$ \nr
 (45,404) | $\psi_{19,175}$ | $\psi_{1,9}$| $-395/101$&$\approx -3.91$ \nr
 (48,413) | $\psi_{19,167}$ | $\psi_{5,43}$| $-505/118$&$\approx -4.28$ \nr
 (55,434) | $\psi_{14,111}$ | $\psi_{9,71}$| $-1699/341$&$\approx -4.98$
\nr
 (54,437) | $\psi_{23,188}$ | $\psi_{11,89}$| $-18815/3933$&$\approx -4.78$
\nr
(59,486) | $\psi_{6,47}$ | $\psi_{5,41}$| $-22201/4779$&$\approx -4.65$
\nr
(62,489) | $\psi_{13,103}$ | $\psi_{9,71}$| $-25195/5053$&$\approx -4.99$
\nr
(61,497) | $\psi_{21,169}$ | $\psi_{7,57}$| $-20485/4331$&$\approx -4.73$
\endtable
\vskip .5cm

\appendix{B}{Solutions of constant energy flux condition for $q<500$}
\vskip .5cm
\begintable
 (p,q) | $\psi$ | $\phi$ | $4 \dpsi +1$ & \cr
 (5,32) | $\psi_{1,9}$ | $\psi_{1,7}$| $-5/2$&$\approx -2.50$ \nr
 (6,37) | $\psi_{1,8}$ | $\psi_{1,7}$ | $-103/37$&$\approx -2.78 $ \nr
 (8,47) | $\psi_{3,17}$ | $\psi_{5,29}$| $-140/47$&$ \approx -2.98$\nr
 (9,58) | $\psi_{4,23}$ | $\psi_{7,45}$| $-209/87$&$\approx -2.40$ \nr
 (10,59) | $\psi_{1,6}$ | $\psi_{1,5}$| $-181/59$&$\approx -3.07$ \nr
 (10,63) | $\psi_{1,4}$ | $\psi_{1,7}$| $-55/21$&$\approx -2.62$ \nr
 (16,101) | $\psi_{5,34}$ | $\psi_{3,19}$| $-511/202$&$\approx -2.53$ \nr
 (16,105) | $\psi_{7,49}$ | $\psi_{9,59}$| $-16/7 $&$\approx -2.29$ \nr
 (18,109) | $\psi_{7,44}$ | $\psi_{13,79}$| $-913/327$&$\approx -2.79$ \nr
 (22,147) | $\psi_{1,10}$ | $\psi_{1,7}$| $-107/49$&$\approx -2.18$ \nr
 (25,166) | $\psi_{9,63}$ | $\psi_{11,73}$| $-917/415$&$\approx -2.21$ \nr
 (32,187) | $\psi_{9,53}$ | $\psi_{7,41}$| $-1117/374$&$\approx -2.99$ \nr
 (31,192) | $\psi_{15,95}$ | $\psi_{5,31}$| $-82/31$&$\approx -2.65$
\nr
 (37,216) | $\psi_{8,47}$ | $\psi_{7,41}$| $-997/333 $&$\approx -2.99$\nr
 (36,241) | $\psi_{11,77}$ | $\psi_{13,87}$| $-1559/723$&$\approx -2.16$ \nr
 (37,248) | $\psi_{8,57}$ | $\psi_{7,47}$| $-2465/1147$&$\approx -2.15$ \nr
 (40,261) | $\psi_{19,121}$ | $\psi_{21,137}$| $-202/87$&$\approx -2.32$ \nr
 (44,269) | $\psi_{19,118}$ | $\psi_{9,55}$| $-8057/2959$&$\approx -2.72$ \nr
 (46,271) | $\psi_{17,101}$ | $\psi_{9,53}$| $-18319/6233 $&$\approx -2.94$\nr
 (45,274) | $\psi_{14,87}$ | $\psi_{11,67}$| $-1129/411$&$\approx -2.75$ \nr
 (47,274) | $\psi_{17,99}$ | $\psi_{29,169}$| $-19313/6439$&$\approx -3.00$
\endtable
\vskip 1cm
\begintable
 (p,q) | $\psi$ | $\phi$ | $4 \dpsi +1$ & \cr
 (51,320) | $\psi_{17,109}$ | $\psi_{11,69}$| $-349/136 $&$\approx -2.57$\nr
 (49,330) | $\psi_{13,91}$ | $\psi_{15,101}$| $-163/77$&$\approx -2.12$ \nr
 (58,339) | $\psi_{23,134}$ | $\psi_{45,263}$| $-9779/3277$&$\approx -2.98$ \nr
 (59,344) | $\psi_{6,35}$ | $\psi_{5,29}$| $-7616/2537$&$\approx -3.00$ \nr
 (59,364) | $\psi_{6,35}$ | $\psi_{5,31}$| $-14347/5369$&$\approx -2.67$ \nr
 (63,368) | $\psi_{4,23}$ | $\psi_{7,41}$| $-1444/483 $&$\approx -2.99$\nr
 (61,372) | $\psi_{8,47}$ | $\psi_{11,67}$| $-5179/1891$&$\approx -2.74$ \nr
 (62,377) | $\psi_{9,53}$ | $\psi_{13,79}$| $-2477/899$&$\approx -2.76$ \nr
 (62,385) | $\psi_{15,91}$ | $\psi_{5,31}$| $-6277/2387$&$\approx -2.63$ \nr
 (65,391) | $\psi_{31,185}$ | $\psi_{61,367}$| $-623/221$&$\approx -2.82$ \nr
 (66,395) | $\psi_{13,79}$ | $\psi_{25,149}$| $-2531/869$&$\approx -2.91$ \nr
 (67,396) | $\psi_{22,131}$ | $\psi_{11,65}$| $-587/201$&$\approx -2.92$ \nr
 (69,410) | $\psi_{32,189}$ | $\psi_{17,101}$| $-2725/943 $&$\approx -2.89$\nr
 (70,411) | $\psi_{21,124}$ | $\psi_{39,229}$| $-2837/959$&$\approx -2.96$ \nr
 (70,419) | $\psi_{11,67}$ | $\psi_{21,125}$| $-8591/2933$&$\approx -2.93$ \nr
 (72,431) | $\psi_{19,115}$ | $\psi_{37,221}$| $-1244/431$&$\approx -2.89$ \nr
 (64,433) | $\psi_{15,105}$ | $\psi_{17,115}$| $-1807/866$&$\approx -2.09$ \nr
 (64,435) | $\psi_{7,44}$ | $\psi_{9,61}$| $-239/116 $&$\approx -2.06$\nr
 (74,439) | $\psi_{31,185}$ | $\psi_{15,89}$| $-47089/16243$&$\approx -2.90$
\nr
 (68,453) | $\psi_{11,70}$ | $\psi_{17,113}$| $-5641/2567$&$\approx -2.20$
\endtable
\vfill

\listrefs
\end